\documentclass[aps,prl,twocolumn,showpacs,amsmath,amssymb,groupedaddress]{revtex4}

\usepackage{graphicx}
\usepackage{dcolumn}
\usepackage{bm}
\begin{document}


\title{Coherent spinor dynamics in a spin-1 Bose condensate}

\author{Ming-Shien Chang}
\author{Qishu Qin}
\author{Wenxian Zhang}
\author{Li You}
\author{Michael S. Chapman}
\affiliation{School of Physics, Georgia Institute of Technology, \\
Atlanta, Georgia 30332-0430, USA}
\date{\today}

\begin{abstract}

Collisions in a thermal gas are perceived as random or incoherent
as a consequence of the large numbers of initial and final quantum
states accessible to the system.  In a quantum gas, e.g. a
Bose-Einstein condensate or a degenerate Fermi gas, the phase
space accessible to low energy collisions is so restricted that
collisions become coherent and reversible.  Here, we report the
observation of coherent spin-changing collisions in a gas of
spin-1 bosons.  Starting with condensates occupying two spin
states, a condensate in the third spin state is coherently and
reversibly created by atomic collisions.  The observed dynamics
are analogous to Josephson oscillations in weakly connected
superconductors and represent a type of matter-wave four-wave
mixing.  The spin-dependent scattering length is determined from
these oscillations to be -1.45(32) Bohr.  Finally, we demonstrate
coherent control of the evolution of the system by applying
differential phase shifts to the spin states using magnetic
fields.

\end{abstract}


\maketitle

Bose-Einstein condensation (BEC) is a well-known phenomenon in
which identical bosons occupy the same quantum state below a
certain critical temperature.  A hallmark of BEC is the coherence
between particles---every particle shares the same quantum wave
function and phase.  Although textbook discussions of
Bose-Einstein condensation typically focus on non-interacting
(ideal) particles, elastic collisions are essential in order for a
quantum degenerate gas to equilibrate.  The inclusion of
collisions also modifies the quantum ground state of the gas,
although it does not change the nature of the coherence of the
condensate---indeed it has been pointed out that collisional
interactions are in fact required to keep the condensate from
fragmenting into multiple nearby quantum states \cite{Nozieres82}.

Collisional coherence is an important theme in quantum degenerate
gases.  For single component condensates, such as spin-polarized
atomic condensates confined in a magnetic trap, the coherence of
the collisional interactions manifests through the coherence of
the many-particle wave function itself, which has been
well-established in early measurements of condensate mean-field
energy \cite{Cornell96Energy} and correlations
\cite{Cornell97coheCorr} as well as in demonstrations of
matter-wave interference \cite{Kett97Interfere} and superfluid
behaviour \cite{Dalibard00Vort, Cornell99Vort, Kett01Vort}.  More
recently, collisional coherence in more complicated systems has
led to remarkable demonstrations including reversible
atom-molecule formation across a Feshbach resonance for both
bosonic and fermionic \cite{Wieman02AMCohere, Jin03MBEC} atoms,
and coherent collisions in optical lattices \cite{Bloch02Mott,
Kasev98Tunnel}.

\begin{figure}
\includegraphics{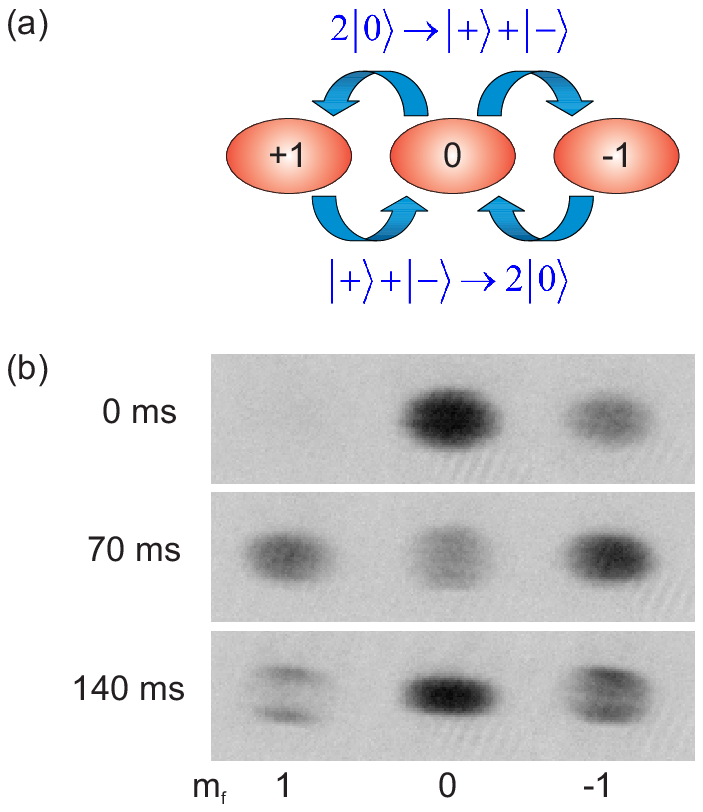}
\includegraphics{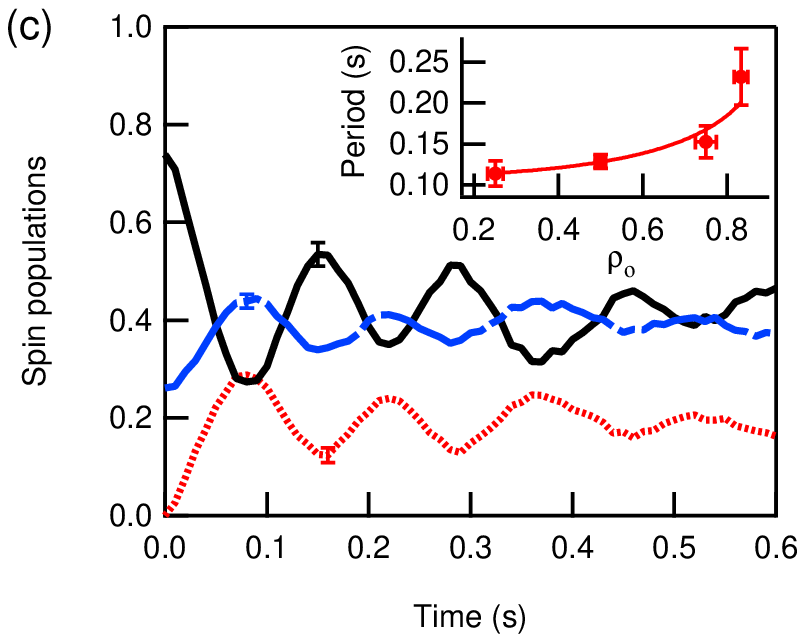}
\caption{\label{fig:Osc31} Coherent spin mixing of spin-1 Bose
condensate in an optical trap. Coherent spin mixing producing
oscillations in the populations of the $F = 1$, $m_F  = 0, \pm 1$
spin states of $^{87}$Rb condensates confined in an optical trap
starting from a superposition of condensate spin components at $t
= 0$ that is subsequently allowed to evolve freely.  a) Schematic
indicates fundamental spin mixing process. b) Absorptive images of
the condensates for different evolution times.  In this example,
the initial relative populations are $\rho _{(1,0,-1)}\simeq
(0,3/4,1/4)$. The condensates are probed 18 ms after release from
the trap, and, to separate the spin components for imaging, a weak
magnetic field gradient is applied for 3 ms during expansion of
the condensates. The field of view is 600 $\mu$m $\times$ 180
$\mu$m. c) Spin populations vs.\ evolution time for the same
initial population configuration showing four clear oscillations.
The damping of the oscillations is due to the breakdown of the
single mode approximation readily apparent in the $t=140$ ms
absorptive image. Here the dotted, solid, and dot-dashed lines
represent the populations in $m_F = 1$, 0 and -1 states.  Inset
shows the measured oscillation period versus the initial
population of the 0 state for different initial superpositions of
$m_F = 0$, -1 states, which compares well with the theoretical
prediction \cite{Pu99}. The (typical) error bars shown are the
standard deviation of three repeated measurements.}
\end{figure}

In this work, we show that the collisional coherence extends to
the internal spin degrees of freedom of a spin-1 Bose gas by
observing coherent and reversible spin-changing collisions. Atomic
Bose condensates with internal spin, so-called spinor condensates
\cite{Ho98, OhmiMachida98, Cornell98phase, Kett98SpinDomain,
Chap01OBEC, Kett03Spin2Trap, Sengstock04, Chap04Spin1,
Ueda00SpinEigenState}, in some cases, are predicted to coherently
inter-convert in a process known as spin mixing, driven solely by
internal interactions in the system  \cite{Bigelow98, Mey99, Pu99,
Tripp04, Rom04, You05CoherentSpin}. In a spin-1 condensate, two
atoms with spin components -1 and +1 can coherently and reversibly
scatter into final states containing two atoms with spin component
0 and vice-versa (Fig.\ \ref{fig:Osc31}a). We observe this process
in a gas of spin-1 $^{87}$Rb bosons confined in an all-optical
trap.  The coherent spin mixing leads to oscillations of the spin
populations, from which we determine the spin-dependent
interaction strength.   This is the first direct measurement of
this important quantity.  The observed spin-mixing is an internal
state analogue to Josephson oscillations in weakly connected
superconductors \cite{Barone82}, and exploiting this analogy, we
demonstrate control of the coherent spinor dynamics using phase
and population engineering. Finally, we use this technique to
drive the spinor condensate to and away from its spin ground
state, which allows us to measure the spin coherence time
\cite{Bigelow98}.

Stimulated by the seminal theoretical work by Ho \cite{Ho98} and
Ohmi and Machida \cite{OhmiMachida98} and early experiments by the
JILA \cite{Cornell98phase, Cornell98Mix2PhaseSep} and MIT
\cite{Kett98SpinDomain} groups, much study has been done on spinor
condensates. Theoretical work has covered ground state structure
\cite{Ho98, OhmiMachida98, Ueda00SpinEigenState, Bigelow98,
You03Spin1}, coherent spinor dynamics \cite{Bigelow98, Mey99,
Pu99, Tripp04, Rom04, You05CoherentSpin}, rotating spinor
condensates \cite{Kett03CorelessVor} and many other topics.  Spin
mixing has been observed in both $F = 1$ and $F = 2$ condensates
\cite{Kett98SpinDomain, Sengstock04, Chap04Spin1, Hirano04Spin2},
although the coherence of this process has not yet been
demonstrated conclusively. Observations thus far have revealed
mostly incoherent relaxation of initially non-equilibrium spin
populations to lower energy configurations from which the sign of
the spin interaction parameter $c_2$ was determined. Although
over-damped single oscillations in spin populations have been
observed in earlier experiments by us and others, their
interpretation has been limited because the initial spin
configurations in these experiments were metastable, and evolution
from these states is noise-driven \cite{Kett98SpinDomain,
Sengstock04, Chap04Spin1, Hirano04Spin2}.  Nonetheless, from these
observations, as well as studies of spin domain formation, it was
possible to determine the magnetic nature of the ground states.

At the microscopic level, the interactions in spinor Bose gases
are determined by spin-dependent 2-body collisions.  In the case
of two colliding spin-$F$ identical bosons, the available
collision channels are restricted by symmetry to those with total
spin $F_{tot}=2F,2F - 2, \cdots ,0$, characterized by s-wave
scattering lengths $a_{F_{tot}}$ at low energies.  We focus on the
$F = 1$ case here, and in the framework of mean-field theory, the
interaction energy including spin can be written as $U(r) = \delta
(r)(c_0  + c_2 \vec F_a  \cdot \vec F_b )$, \cite{Ho98,
OhmiMachida98}  where $r$ is the distance between two atoms $a,b$
and $c_0  = 4\pi \hbar ^2 (a_0  + 2a_2 )/3m$ and $c_2  = 4\pi
\hbar ^2 (a_2  - a_0 )/3m$, where $m$ is the atomic mass and
$a_{0,2} $ are the s-wave scattering lengths for total spin 0,2
channels.  For $^{87}$Rb atoms in the $F = 1$ hyperfine state, the
scattering lengths the $a_{0,2} $ are nearly equal, and hence the
spin-dependent mean field energy $c_2 n$ is very small (only 200
pK \cite{Greene01Rb, Verhaar02Rb} for typical densities, $n \sim
10^{14} $ cm$^{-3}$), compared to both the scalar mean field $c_0
n$, and the estimated temperature of the gas, ~50 nK. Nonetheless,
the small spin-dependent mean field couplings are non-negligible
and lead to qualitatively different ground state structures
depending on the sign of $c_2 $, being ferromagnetic ($c_2  < 0$)
for $^{87}$Rb  \cite{Sengstock04, Chap04Spin1, Greene01Rb,
Verhaar02Rb} or anti-ferromagnetic ($c_2  > 0$) for $^{23}$Na
 \cite{Kett98SpinDomain, Bohn98}.  Moreover, these spinor
interactions yield a rich variety of coherent and incoherent
phenomena including coherent spinor mixing, spin squeezing and
entanglement \cite{Bigelow98, Zoller02EntangleSpin}, spin domain
formation, and spinor vortices.

A single-component (scalar) atomic condensate with a large number
of atoms is well-described within a mean-field treatment by an
order parameter (`condensate wave function') governed by nonlinear
Schr\"{o}dinger or Gross-Pitaevskii (G-P) equation.  For an $F =
1$ spinor condensate, the three spin components $m_F  = 1,0, - 1$
are described by a vector order parameter, $\vec \psi (\vec r,t) =
(\psi _1 ,\psi _0 ,\psi _{ - 1} )$ which is governed by a set of
three coupled G-P equations \cite{Ho98, Bigelow98}
\begin{eqnarray}
i\hbar \frac{{\partial \psi _1 }}{{\partial t}} &=& L_1 \psi _1  +
c_2 (n_0 + n_1  - n_{ - 1} )\psi _1  + c_2 \psi _{ - 1}^* \psi
_0^2 \\
i\hbar \frac{{\partial \psi _0 }}{{\partial t}} &=& L_0 \psi _0 +
c_2 (n_1  + n_{ - 1} )\psi _0  + 2c_2 \psi _0^* \psi _1 \psi _{ -
1} \\
i\hbar \frac{{\partial \psi _{ - 1} }}{{\partial t}} &=& L_{ - 1}
\psi _{ - 1}  + c_2 (n_0  + n_{ - 1} - n_1 )\psi _{ - 1} + c_2
\psi _{ + 1}^* \psi _0^2 \nonumber \\
& &
\end{eqnarray}
where $L_{ \pm 1,0}  = ( - \hbar ^2 \nabla ^2 /2m + V_t + E_{\pm
1,0}  + c_0 n - \mu )$, $V_t $, $E_{\pm 1,0} $ and $n_{ \pm 1,0} $
are the optical trapping potential, Zeeman energies and densities
for corresponding Zeeman projections, $\mu $ is the chemical
potential, and $n = n_1  + n_0 + n_{-1} $ is the total density.
The coherent spin mixing of the internal populations responsible
for oscillations of the spin populations are determined by the
last term in Eqs.1-3.  However, the interplay of this process with
the particle exchange collisions, represented by the penultimate
term in Eqs.\ 1-3 poses challenging problems both in theory and in
experiment because they occur with the same time scale (typically
$<$ 10 Hz for $^{87}$Rb), and they are both very sensitive to
external magnetic fields and field gradients represented in $E_{
\pm 1,0} $  \cite{Kett98SpinDomain, Sengstock04,
Ueda00SpinEigenState, You05CoherentSpin, Hirano04Spin2, Bigelow00,
Kett99prl}.  Hence, in general, the dynamics described by these
equations reveal a rich coupling between the internal and external
degrees of freedom of the condensate components resulting in a
variety of observed phenomena including spin mixing
\cite{Kett98SpinDomain, Sengstock04, Chap04Spin1}, spin domain
formation  \cite{Kett98SpinDomain}, and spin textures
 \cite{Kett03CorelessVor, Cornell99SpinUntwist}.

Although the internal and external dynamics are generally
inseparable, under certain conditions they can be decoupled.  In
particular, when the available spin interaction energy is
insufficient to create spatial spin structures in the condensates,
then the external dynamics will be suppressed.  This occurs when
the spin healing length $\xi _s  = h/\sqrt {2m\left| {c_2 }
\right|n} $ is larger than the size of the condensate.  In this
case, then $\psi _1 $, $\psi _0 $ and $\psi _{ - 1} $ share the
same spatial wave function $\sqrt {n(\vec r)} e^{ - i\mu t/\hbar }
$ which allows for a considerable simplification of Eq.\ 1-3. This
is known as the single mode approximation (SMA), with which the
order parameter reduces to $\sqrt {n(\vec r)} e^{ - i\mu t/\hbar }
\vec \chi $, where $\vec \chi ^{\text{T}}  = (\sqrt {\rho _1 }
e^{i\phi _1 } ,\sqrt {\rho _0 } e^{i\phi _0 } ,\sqrt {\rho _{ - 1}
} e^{i\phi _{ - 1} } )$ and $\rho _i $ and $\phi _i $ represent
the fractional population and phase of the $i^{th}$ Zeeman state.
With this approximation, the internal dynamics take on a
particularly simple form determined by just two dynamical
variables, $\rho _0 (t)$, the fractional population of the 0
state, and the relative phase of the spinor components $\phi (t)
\equiv \phi _{1} + \phi _{-1} - 2\phi _0 $  \cite{Pu99}.  The
populations of the other states are directly determined by $\rho
_{ \pm 1}  = (1 - \rho _0 \pm M)/2$, where  $M = (N_{1}  - N_{-1}
)/N$ is the global magnetism which is a constant of the motion,
and $N_i $ is the number of atoms in the $i^{\text{th}}$ Zeeman
projection with $N = N_1 + N_0 + N_{-1}$.  The energy
(Hamiltonian) of the system in a uniform magnetic field then takes
the following simple form  \cite{Rom04, You05CoherentSpin}:
\begin{equation} E = c\rho _0 [(1 - \rho _0  +
\sqrt {(1 - \rho _0 )^2  - M^2 } \cos \phi )] + \delta (1 - \rho
_0 ),
\end{equation}
where $c = c_2 N\int {\left| {\psi (r)} \right|^4 dr} $ is the
effective spin-mixing rate, and $\delta = (E_{1} + E_{-1} - 2E_0
)/2 \simeq 2\pi \hbar  \cdot $ ($72B^2 $ Hz) is the difference in
energies of the magnetic Zeeman levels in a field of $B$ gauss.
Within the SMA, the ground state spin populations and relative
phase are readily found for arbitrary magnetization and magnetic
field by minimizing Eq.\ 4.  In particular, for $c(c_2 ) < 0$, the
energy of the system is minimized at low fields for relative
phases $\phi  = 0$ and population $\rho _0  = (1 - M^2 )/2$
 \cite{Pu99, You05CoherentSpin}.  For other non-equilibrium
populations or phases, the system will have excess spin energy
that can drive a coherent evolution of the spinor system.

The evolution of the system will follow Hamiltonian equations of
motion derived from Eq.\ 4  \cite{You05CoherentSpin}:
\begin{eqnarray}
\dot \rho _0 &=& \frac{{2c}}{\hbar }\rho _0 \sqrt {(1 - \rho _0
)^2 - M^2 } \sin \phi \\
\dot \phi  &=&  - \frac{{2\delta }}{\hbar } + \frac{{2c}}{\hbar
}(1 - 2\rho _0 ) \nonumber \\
& & + \frac{{2c}}{\hbar }\frac{{(1 - \rho _0 )(1 - 2\rho _0 ) -
M^2 }}{{\sqrt {(1 - \rho _0 )^2  - M^2 } }}\cos \phi .
\end{eqnarray} These coupled equations are nonlinear
Josephson-type equations and point to the equivalency of spin
mixing in a spin-1 condensate to Josephson systems realized in
superconductors \cite{Barone82} and other superfluids
 \cite{Kasev98Tunnel, Cornell98phase, Leggett75, Wheatley75,
Pack01He3, Javan86, Shenoy99, Inguscio01JOE}.  The non-linearity
of these equations provides a rich manifold of dynamical
trajectories that can be accessed experimentally by choice of
initial populations and phases of the spin components and the
strength of the applied magnetic field.

To investigate the coherent dynamics of this system, we begin with
$^{87}$Rb condensates created using an improved version of the
all-optical trapping technique we have previously reported
\cite{Chap01OBEC, Chap04Spin1}.  Using a dynamical compression
technique and just a single focused laser beam, condensates with
up to $300,000$ atoms are created after 2 s of evaporative
cooling. The condensates created in this optical trap are
generally in a mixture of all $F = 1$ spin states and reveal
complicated spatial domains.  To create a well-characterized
initial condition, we first prepare a condensate in the $\left| {F
= 1,m_F  =  - 1} \right\rangle $ state by applying a magnetic
field gradient during the evaporative cooling.

To initiate spin dynamics, a coherent superposition of spin states
with non-equilibrium spin populations is created by applying a
sequence of phase-coherent microwave pulses tuned to $F = 1
\leftrightarrow F = 2$ transitions. Following this state
preparation, the condensate is allowed to freely evolve in the
optical trap.  A typical evolution is shown in Fig.\
\ref{fig:Osc31}c for an initial spin configuration of $\rho
_{(1,0, - 1)}  \simeq (0,3/4,1/4)$. Up to four distinct
oscillations are observed in this example before the spin
populations damp to a steady state. These oscillations demonstrate
the coherence of the spin mixing process.

\begin{figure}
\includegraphics{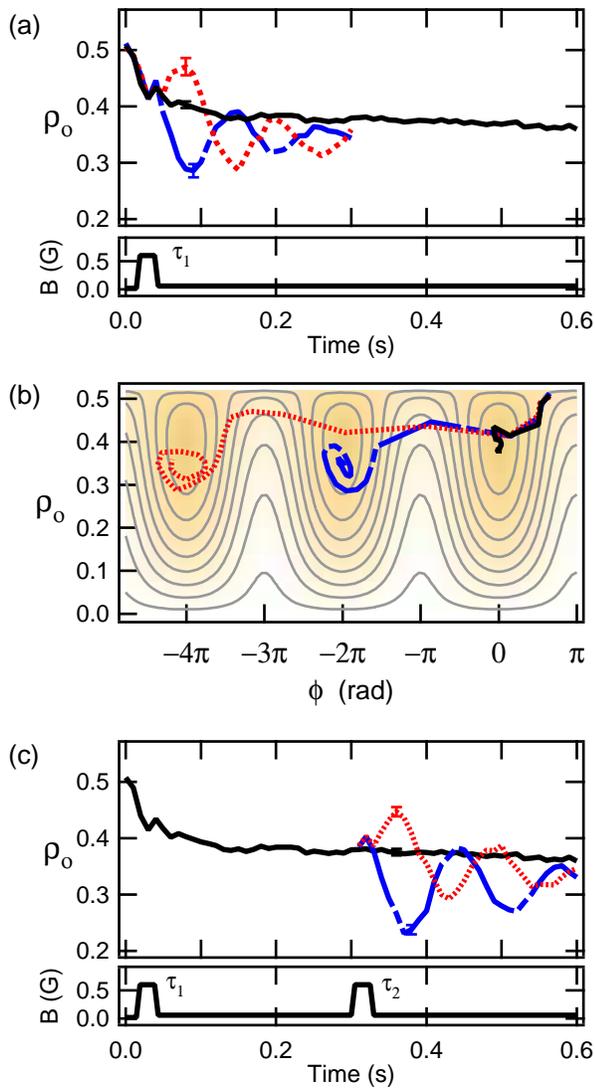}
\caption{\label{fig:CtrlCoherentOsc} Coherent control of spinor
dynamics. a) An initial spin configuration of $\rho _{(1,0,-1)}
\simeq (0,1/2,1/2)$ is allowed to evolve in a field of 15 mG for
14 ms at which point the populations reach the values
corresponding to the ferromagnetic ground state at this
magnetization: $\rho _{(1,0,-1)} \simeq (1/16,3/8,9/16)$. Then, a
pulse of 600 mG field is applied to shift the spinor phase.  The
dashed, solid and dotted curves represent pulse widths of $\tau _1
=20$, 24.4 and 30 ms respectively. For certain applied phase
shifts, the coherent spin mixing can be halted. This occurs for
$\tau _1 = 24.4$ ms corresponding to phase shift $\Delta \phi =
-2.5\pi $ and for $\tau _1 = 5.3$ ms corresponding to $\Delta \phi
= -0.5\pi $. b) Reconstructed dynamical trajectories of the system
determined by fitting the experiment data to Eq.\ 3,4 including a
phenomenological phase damping term.  The free parameters of the
fit are the damping coefficient and the unknown (but reproducible)
initial spinor phase resulting from the state preparation that
depends on the applied microwave pulse width and the duration in
the upper hyperfine manifold.  The contours show curves of equal
energy.  c) To investigate the spin coherence of the ground state
spinor created by the first pulse with $\tau _1 = 24.4$ ms, a
second pulse is applied at 300 ms to reestablish the oscillations.
The solid, dashed and dotted curves corresponds to $\tau _2 = 0$,
10 and 20 ms respectively.  The typical error bars shown are the
standard deviation of three repeated measurements.}
\end{figure}

We have measured the spin oscillation frequency for different
initial spin populations.  These data are shown in the inset of
Fig.\ \ref{fig:Osc31}c. and show good agreement with theoretical
predictions, $c\sqrt {1 - \rho _0^2 } $  \cite{Pu99,
You05CoherentSpin} which can be derived from Eq.\ 5-6.  These
measurements provide a direct determination of the magnitude of
the spin interaction energy \cite{Bigelow98}, $\left| c
\right|/\hbar  = 2\pi  \times 4.3(3)$ rad/s for our system.  These
oscillation frequencies at low magnetic field do not determine the
sign of $c(c_2 )$, however it was established by previous studies
of the nature of the ground state (and confirmed in the present
study by the results shown later in Fig.\
\ref{fig:CtrlCoherentOsc})  that $c_2  < 0$ for the F = 1 manifold
of $^{87}$Rb. This value of $c$, combined with the measured
condensate density, $n = 2.1(4) \times 10^{14} $ cm$^{-3}$,
determined from the rate of the condensate expansion during
time-of-flight, permits determination of $c_2 $, or equivalently,
the difference in scattering lengths $a_2 - a_0  = -1.45(32)a_B $,
where Bohr radius $a_B = 0.529$ \AA. This is the first direct
measurement of this important quantity, and our value agrees with
the theoretical determination of $a_2 - a_0 = -1.40(22)a_B $
derived from photoassociative spectroscopic and Feshbach resonance
data \cite{Greene01Rb, Verhaar02Rb}.

The oscillations are observed to damp with a time constant of 250
ms, and the damping coincides with the appearance of spatial spin
structures apparent in the images in Fig.\ \ref{fig:Osc31}b. These
structures indicate the invalidation of the SMA underlying Eqs.\
5-6 and lead to a complicated interplay of the internal and
external dynamics that ultimately transfers the internal spin
energy into spatial domain structures \cite{Pu99}.  A detailed
study of these structures will be the focus of future work.

\begin{figure}
\includegraphics{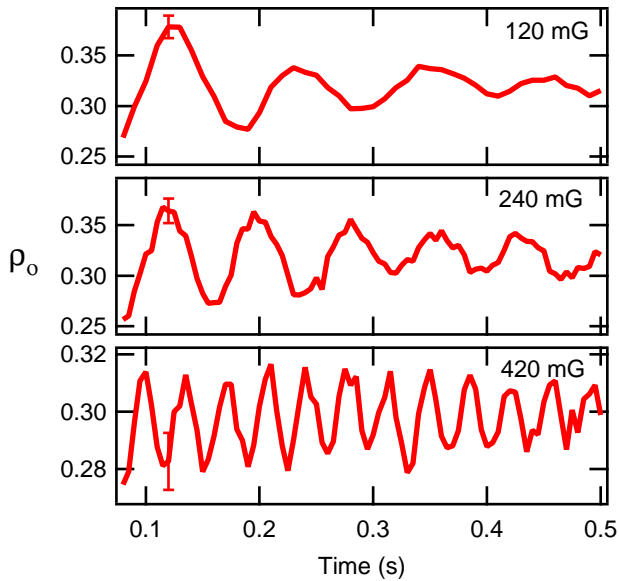}
  \caption{\label{fig:OscVsBfield}
Coherent spin mixing vs.\ magnetic field.  An initial
non-equilibrium spin population configuration of $\rho _{(1,0,
-1)} \simeq (0,1/2,1/2)$ is created and allowed to evolve in a
field of 15 mG for 70 ms to allow for maximum spin mixing.  At
this point, the magnetic field is ramped to different levels.
Subsequently, the system displays small amplitude oscillations
analogous to the AC Josephson effect, $\rho _{0}(t)\propto \delta
^{-1} \sin 2\delta t$. The typical error bars shown are the
standard deviation of three repeated measurements.}
\end{figure}

The large amplitude oscillations observed in Fig.\
\ref{fig:Osc31}c are in the nonlinear regime of Eqs 5-6.  It is
also possible to access the linear regime more typical of the
standard Josephson effect by tuning the parameters of the system.
In particular, for large applied magnetic fields such that $\delta
\gg c$ and appropriate initial populations, the phase evolution is
dominated by the quadratic Zeeman effect of the external field,
and Eq.\ 6 reduces to $\dot \phi \simeq -2\delta /\hbar $ . For
these conditions, the system exhibits small oscillations analogous
to AC-Josephson oscillations, $\rho _0 (t) \simeq A\delta ^{-1}
\sin 2\delta t$, where $A$ is determined by the initial
populations. We have observed these oscillations as shown in Fig.\
\ref{fig:OscVsBfield}. for different applied magnetic fields. Up
to 12 fast oscillations are observed at the highest fields that
were studied, where the time scale of the internal spinor dynamics
is better separated from the time scale for the formation of
spatial spin structures. The frequency of the measured
oscillations vs.\ the magnetic field matches within 10\% of the
prediction $\Omega = 2\delta $, while the $\delta ^{-1}$ scaling
for the amplitude is seen only for higher fields, presumably due
to the invalidity of the SMA for larger amplitude oscillations. In
the future, being able to tune the system to the linear regime
provides a path to study many analogous effects previously
observed in Josephson systems such as Shapiro levels
\cite{Barone82, Bigelow00, Pack01He3, Shapiro63} by including a
time-varying component to the applied magnetic field.

Beyond controlling the system via the initial conditions, the
dynamical evolution of the system can be controlled in real time
by either changing spin populations and/or changing the spinor
phase $\phi $.    We demonstrate that we can coherently control
the dynamical evolution of the spinor by applying phase shifts,
and in particular we drive the systems to the ferromagnetic spinor
ground state using this technique.  In this experiment, an initial
non-equilibrium spin configuration is created and allowed to
evolve for a fraction of an oscillation until $\rho _0 (t)$
reaches the ground-state ratio $\rho _{0,gs}=(1-M^2)/2$
\cite{Pu99, You03Spin1}.  At this point, the system is not in the
ground state because $\phi \ne (\phi _{gs}=0)$ (and it is still
oscillating!)  At this moment, we briefly pulse on a magnetic
field of 0.6 G to apply a phase shift to the spinor, $\Delta \phi
= \int {\delta (t)dt}.$  The evolution of the system is recorded
in Fig.\ \ref{fig:CtrlCoherentOsc}a for different pulse durations.
We find that for particular applied phase shifts, the spinor
condensate is brought to its ground state, evidenced by the
subsequent lack of population oscillation. For other applied phase
shifts, the system is driven to different points in the phase
space of the system, for which the subsequent evolution of the
system is dramatically different and exhibits oscillations.

It is possible to reconstruct the dynamical trajectories of the
system using the measured $\rho _{0}(t),$ along with the known
applied phase shifts and the equations of motion, Eqs.\ 5-6.
Although the damping evident in the measurement is due to the
spatial dynamics coupled to the internal spin mixing dynamics,  a
phenomenological phase damping term may be added to Eq.\ 6 to
represent the spatial varying spin mixing rate which is
responsible for damping the population oscillation.  The
reconstructed trajectories show good qualitative agreement with
the measurements in the time domain and are plotted on the phase
space diagram of the system (Fig.\ \ref{fig:CtrlCoherentOsc}b).
Also shown in the figure are the contours of equal energy of the
spinor given by Eq.\ 4. The trajectories clearly show that the
system tends to damp to the minimum energy points (i.e. the
ferromagnetic spinor ground state), which is $\rho _0  \simeq
3/8$, $\phi _{gs} = 0\;\bmod (2\pi )$ for $M = 1/2$, $c<0$ and
$\delta \approx 0$.  For the case of anti-ferromagnetic
interactions, such as in $^{23}$Na, $c > 0$, the energy contours
differ only in sign, and the system would instead relax to $\rho
_0 =1$, $\phi _{gs} = \pi \;\bmod (2\pi )$ within the validity of
the SMA  \cite{Pu99, You03Spin1}.

To demonstrate explicitly the coherence of the spinor ground
state, we impart a second phase shift to the system at later times
to displace the system to a different point in phase space.  As
anticipated, the second phase shift is found to re-initiate the
spin mixing dynamics (Fig.\ \ref{fig:CtrlCoherentOsc}c) when
$\Delta \phi  \ne 0\;\bmod (2\pi )$.  We have used this technique
to determine the ground state spinor decoherence time by measuring
the amplitude of the subsequent oscillations for different delay
times of the second pulse.  The spinor decoherence time is found
to be 3 s, which is approximately the lifetime of the condensate
and is much longer than the damping time of spin population
oscillations.

As noted, the damping of the spin oscillations coincides with the
appearance of spin wave-like spatial structures in the spinor wave
function (see images in Fig.\ \ref{fig:Osc31}b).  Hence it is
clear that the SMA is not strictly valid for our system.  These
waves derive their energy from the internal (spin) degrees of
freedom, and it is this energy transfer that ultimately damps the
spin mixing.  On the other hand, if the spinor condensate is
driven to its ferromagnetic ground state, as shown in Fig.\
\ref{fig:CtrlCoherentOsc}b, there is no internal (spin) energy
available for the motional degrees of freedom, and spatial spin
structures cannot form. Indeed in this case, the three spin
components are observed to have the same spatial wave function and
appear to be miscible.

The observation of coherent spinor dynamics in a ferromagnetic
spin-1 system reported here paves the way for a host of future
explorations. These systems are predicted to manifest complex
quantum correlated states exhibiting entanglement and squeezing,
and in general, it will be very interesting to explore the regime
of small atom number $ < 1000$, where sub-shot noise effects
should become important \cite{Bigelow98}.  Viewing the spin-mixing
dynamics as a type of internal Josephson effect, many future
explorations and manipulations of the system can be envisaged
following along the path of superconducting weak-links. Finally,
the coupling of the internal dynamics to the spatial wave function
can be avoided in future experiments by either decreasing the
condensate radii relative to the spin healing length $\xi _s $
and/or operating at high magnetic fields where the time scales for
mixing and damping are better separated.  On the other hand, the
coupling of the internal and external degrees of freedom in this
system provide a new system for exploring nonlinear atom optical
phenomena such as spatial-temporal dynamics of four-wave mixing
 \cite{Phil98}.

We note that a group in Mainz, Germany has independently observed
coherent spin-mixing oscillations in a Mott state of atoms on a
lattice  \cite{Bloch05CoherentSpin}.  Their experiment involves a
system of many copies of two atoms in each lattice site.  On the
other hand, our system involves a few hundred thousand atoms and
the observed coherence reflects the presence of macroscopic
quantum fields.

\subsubsection{Methods}

To create the condensates, we begin by collecting up to $5 \times
10^8 $ cold atoms in a simple vapour cell $^{87}$Rb
magneto-optical trap (MOT), which is overlapped with an optical
trap formed by a single CO$_2$ laser beam of 70 W focused to a
waist of 60 $\mu$m.  Up to $3.7 \times 10^7 $ atoms are loaded
into the optical trap, at a density of $4 \times 10^{13} $
cm$^{-3}$.  In order to achieve higher densities for efficient
evaporation, the trap is compressed immediately after loading by
smoothly changing the laser focus to 26 $\mu$m over 600 ms using a
zoom lens. Simultaneously, the laser power is ramped down over 1.8
s to lower the trap depth and force evaporative cooling. Using
this technique, mostly pure condensates containing up to $3 \times
10^5 $ atoms are created.  This technique is not only simple and
fast, but also produces condensates 10 times larger than our
previous methods  \cite{Chap01OBEC, Chap04Spin1}.  The condensates
created in this optical trap are generally in a mixture of all $F
= 1$ spin states and reveal complicated spatial domains.  To
create a well-characterized initial condition, we apply a magnetic
field gradient during the evaporative cooling \cite{Chap04Spin1},
which results in pure $m_F = -1$ condensate containing 150,000
atoms-this state is stable against both local and global spin
dynamics due to the conservation of angular momentum.  The trap
frequencies are 2$\pi$(190,170,17) rad/s, and the condensate
density and the Thomas-Fermi (T-F) radii are estimated to be $2.1
\times 10^{14} $ cm$^{-3}$ and (3.2,3.6,36) $\mu$m respectively.
The lifetime of the condensate is measured to be 3 s.  Coherent
spin state superpositions are created from the pure $m_F  =  - 1$
condensates by applying a sequence of phase-coherent microwave
pulses tuned to $F = 1 \leftrightarrow F = 2$ transitions.  The
pulses are applied at a field of 420 mG to separate out the
transitions between the different Zeeman sub-levels.  Following
the pulse sequence, the magnetic field is ramped from 420 mG to 15
mG in 10 ms.  Typical pulse lengths are 20  s for a $F = 1
\leftrightarrow F = 2$ pulse.

Acknowledgements. This work was supported by NSF-PHYS 0303013 and
NASA-NAG3-2893. We would like to thank C. Hamley, K. Fortier, J.
Sauer and other members of the Georgia Tech Atomic Physics and
Quantum Optics Group for their assistance, and H. Pu for valuable
discussions.

\end{document}